\documentclass[aps,prx,twocolumn,superscriptaddress]{revtex4-2}
\usepackage{amssymb}
\usepackage{amsmath,bm}
\usepackage{graphicx,color}
\usepackage{bbm}
\usepackage{newtxmath}
\usepackage{multirow}
\usepackage{xcolor}
\usepackage{hyperref}

\newcommand{\ah}[1]{\textcolor{black}{#1}}

\newcommand{\mcm}[1]{\textcolor{black}{#1}}

\newcommand{\MM}[1]{\textcolor{black}{#1}}

\newcommand{\be}{\begin{equation}}
\newcommand{\ee}{\end{equation}}
\newcommand{\bea}{\begin{eqnarray}}
\newcommand{\eea}{\end{eqnarray}}

\newcommand{\g}{\mathbf{g}}
\newcommand{\go}{\mathbf{g}_0}

\newcommand{\figref}[1]{\figurename~\ref{#1}}
\newcommand{\appref}[1]{App.~\ref{#1}}	
\newcommand{\Mp}{\mathcal{M}_P}
\newcommand{\Ma}{\mathcal{M}_A}

\begin{document}

\title{Anomalous elasticity of cellular tissue vertex model}

\author{Arthur Hernandez}
\affiliation{Department of Physics, University of California Santa Barbara, Santa Barbara, CA 93106}
\author{Michael F. Staddon}
\affiliation{Center for Systems Biology Dresden, Dresden, Germany}
\affiliation{Max Planck Institute of Molecular Cell Biology and Genetics, Dresden, Germany}
\affiliation{Max Planck Institute for the Physics of Complex Systems, Dresden, Germany}
\author{Mark J. Bowick}
\affiliation{Department of Physics, University of California Santa Barbara, Santa Barbara, CA 93106}
\affiliation{Kavli Institute for Theoretical Physics, University of California Santa Barbara, Santa Barbara, CA 93106}
\author{M. Cristina Marchetti}
\affiliation{Department of Physics, University of California Santa Barbara, Santa Barbara, CA 93106}
\author{Michael Moshe}
\email{michael.moshe@mail.huji.ac.il}
\affiliation{Racah Institute of Physics, The Hebrew University of Jerusalem, Jerusalem, Israel 91904}


\begin{abstract}
 Vertex Models, as used to describe cellular tissue, have an energy controlled by deviations of each cell area and perimeter from  target values. The constrained nonlinear relation between area and perimeter leads to new mechanical response. Here we provide a mean-field treatment of a highly simplified model: a uniform network of regular polygons with no topological rearrangements. Since all polygons deform in the same way, we only need to analyze the ground states and the response to deformations of a single polygon (cell). The model exhibits the known transition between a fluid/compatible state, where the cell can accommodate both target area and perimeter, and a rigid/incompatible state.
We calculate and measure the mechanical resistance to various deformation protocols and discover that at the onset of rigidity, where a single zero-energy ground-state exists, 
linear elasticity fails to describe the mechanical response to even infinitesimal deformations. In particular we identify a breakdown of reciprocity expressed via different moduli for compressive and tensile loads, implying  non-analyticity of the energy functional.
We give a pictorial representation in configuration space that reveals that the complex elastic response of the Vertex Model arises from 
the presence of two distinct sets of reference states (associated with target area and target perimeter).
Our results on the critically compatible tissue provide a new route for the design of mechanical meta-materials that violate or extend classical elasticity. 
\end{abstract}

\maketitle
\section{Introduction}
{  Biological tissue are active materials capable of generating mechanical stresses and transmitting such stresses at the organ and organism scale  ~\cite{gomez2020measuring}. Their ability to tune rigidity and adapt the mechanical response to external perturbations engender significant challenges for the formulation of a continuum mechanics. } 
Of special interest are epithelial tissue - two-dimensional layers of tightly packed cells {that can spontaneously undergo transitions between liquid-like states where cells freely exchange neighbors and solid-like states where cells are jammed~\cite{angelini2011glass,puliafito2012collective,sadati2013collective,park2015unjamming,mongera2018fluid}. Unlike solid-liquid transitions in inert matter or jamming transitions in granular materials, the rigidity transition of confluent tissue occurs at constant density and is driven by two classes of mechanisms: \emph{active processes} and \emph{geometrical constraints}. }
{Active processes, such as cell motility or fluctuations in the tension of the cell-edge network, maintain the tissue out of equilibrium, 
facilitating or impeding cellular rearrangements and strongly altering the fluidity/rigidity of the cell collective~\cite{bi2015density,bi2016motility,krajnc2020solid,krajnc2021active}. 
Geometrical frustration of the cellular network provides a different path to rigidity associated with \emph{geometric incompatibility} and akin to the one found in metamaterials and biopolymer networks~\cite{storm2005nonlinear,chen2018branches}. This geometry-driven transition between rigid and floppy states  has been identified before in Vertex and Voronoi models of confluent tissue~\cite{moshe2018geometric,sahu2019nonlinear,damavandi2021energeticI,damavandi2021energeticII}, {but the characterization of the elastic and rheological response of the VM to external deformations is only beginning to be addressed~\cite{tong2021linear}.} }

The formulation of continuum elastic theories of solids crucially relies on the existence of a potential energy and of a unique reference state. 
Both  are absent in living matter, where  out-of-equilibrium active processes cannot be captured by a conservative potential energy and the under-constrained structure of the cellular network results in degenerate ground states. 
As a result,  the formulation of a continuum elasticity of living matter remains a formidable challenge.

In this paper we examine how  geometric constraints affect the continuum elasticity of cellular networks in the context of a regular  two-dimensional  Vertex Model (VM). The VM describes a confluent tissue as  a network of polygons tiling the plane. Each polygon represents a cell and is characterized by target
values of area and perimeter encoding a variety of bio-mechanical mechanisms \cite{honda1983geometrical,nagai2001dynamic,staple2010mechanics, farhadifar2007influence, chiou2012mechanical,fletcher2014vertex, bi2015density, alt2017vertex, barton2017active}.   {The observed cells area and perimeter are controlled by a tissue energy that penalizes deviations from  target values. \mcm{Many recent studies} of the  VM have focused on disordered and active realizations, \mcm{consisting of a  disordered network of irregular polygons with active processes driving} cell rearrangements and neighbor exchanges~\cite{merkel2017triangles,merkel2018geometrically,popovic2021inferring,grossman2021instabilities}. Here, in contrast, we consider an ordered realization where the network is composed of regular polygons and neglect active processes responsible, for instance, for $T_1$ transitions.  This allows us to isolate the structural and energetic origin of the rigidity transition associated with geometric incompatibility. }

By combining analytical methods and numerical simulations, we show that at the onset of rigidity, \mcm{i.e., at the transition between the compatible and incompatible regimes,} the response of the VM to infinitesimal deformations \mcm{cannot} be described by linear Hookean elasticity.
\mcm{Specifically, at the critical point}  mechanical reciprocity is violated, an anomalous coupling between bulk and shear deformations emerges, and quartic rigidity is observed in response to shear deformations. 
Additionally, the fluid state exhibits vanishing stiffness up to a critical strain as it can accommodate external strains with zero stress by spontaneous shear. \mcm{In contrast, the rigid state has finite linear response that is captured by linear elasticity.}

{Very recent work that involves one of us~\cite{huang2021shear} has examined numerically the response of a \emph{disordered} Voronoi model that naturally incorporates topological rearrangements to  quasi-static shear. This work  also finds that the compatible/fluid state exhibits zero stress below a critical applied strain, confirming the results of our minimal mean-field approach. It additionally shows that both the liquid and the solid exhibit shear stiffening above a critical strain and that a mean-field theory  that incorporates the ground state degeneracy of the compatible regime \mcm{inspired by the one shown in the present paper} captures the nonlinear behavior of the shear response.} 

Although derived from an energy functional, the elasticity of the critically-compatible VM shares similarities with odd elasticity, including the breakdown of reciprocity and the emergence of  an anomalous coupling between isotropic and shear deformations. As in odd solids, the linear response of the VM violates the  basic symmetries of the elastic stiffness tensor of passive solids.
Contrary to odd elasticity, these properties emerge {not from a sustained energy input that breaks the conservative nature of forces,} but from {pure geometric constraints that result in} the failure of a Taylor expansion to faithfully describe the elastic potential energy {even for small deformations}.
{The geometric origin of the anomalous elasticity is highlighted through a generalized continuum elastic theory of the VM and a corresponding pictorial description,}
which provides excellent quantitative agreement with the numerical simulations. 
{Our findings provide new insights into the geometrical aspects of tissue mechanics and emergent rigidity, which underlie in an essential way the rigidity transitions controlled by active processes. They also lay out a path for} the design of new mechanical metamaterials with  mechanical properties mimicking those of living tissue.
 
The structure of this paper is as follows: In Sect.~\ref{sec:VM} we review the properties of the passive ordered VM.  In Sect.~\ref{sec:MF} we introduce a mean-field approach to VM, implemented in Sect.~\ref{sec:response} to measure global response to uniform loads in the compatible and incompatible states, and compare with numeric results. 
At the heart of our work, in Sect.~\ref{sec:Violation} we focus on the critically compatible case and show that the measured properties violate linear elasticity due to ill defined elastic constants. Sect.~\ref{sec:Visual} proposes a visual representation of VM mechanics, uncovering the source of its peculiar behavior and the required modifications to classical elasticity. The last section \ref{sec:summary} provides a brief summary and offers directions  for the road ahead.

\section{Vertex Model and geometric incompatibility}
\label{sec:VM}
In the VM, each cell is described as a convex polygon with {target} area and perimeter $A_0$ and  $P_0$, respectively. Given a configuration with actual area $A$ and perimeter $P$, the cell stores a mechanical energy   
\begin{equation}
	E_\mathrm{cell}=\frac{\kappa_A}{2}\big(A-A_0\big)^2 +\frac{\kappa_P}{2}\big(P-P_0\big)^2\;.
	\label{eq:Ea}
\end{equation}
A confluent tissue consists of a network  of many such cells, covering the plane. 
Cells in epithelial tissue typically resemble disordered arrangements of mainly $5-$, $6-$ and $7-$sided irregular polygons, with an average coordination number of $3$ at each vertex. 
To highlight the mechanics emerging from purely geometric constraints, here
 we consider the seemingly simple  quasi-static response of a uniform tissue (i.e., uniform $A_0$ and  $P_0$) to uniform imposed loads {that lead to uniform observed $A$ and $P$.} We  assume that all cells respond identically. {Thus the tissue energy} is $E = N\, E_\mathrm{cell}$ and it is sufficient to analyze the behavior of a single cell. {This corresponds to a mean-field theory of the tissue VM, where spatial variations are either irrelevant or negligible.} 
Additionally, for clarity we mainly analyze the case of a triangular tissue. This transparent example is also closer to the familiar discrete model of elastic materials \cite{seung1988defects,lubensky2015phonons,sheinman2012nonlinear}. {Our results are not qualitatively affected by the specific polygonal shape considered when uniform remote loads \mcm{resulting in affine deformations} are imposed.} 
A na\"ive degree-of-freedom counting reveals that the VM is under-constrained~\cite{yan2019multicellular,damavandi2021energeticI}. Even the most rigid polygon, a triangular cell, has three structural degrees of freedom, corresponding to the lengths of the three edges, but {fixing target area and perimeter only imposes  two constraints}, implying that a single triangular cell is floppy.

Recent work has shown that VMs exhibit a transition tuned by the \emph{target} shape parameter $s_0=P_0/\sqrt{A_0}$ between a fluid-like state where cells freely intercalate and a rigid state where cells are collectively jammed~\cite{bi2015density,bi2016motility,park2015unjamming}. {The order parameter for this transition is the \emph{observed} cell shape defined as $s=P/\sqrt{A}$.} 
In early work the loss of rigidity was associated with the vanishing of energy barriers for neighbor exchanges known as $T_1$ transitions which mediate local changes in network topology~\cite{bi2015density}. Studies of VMs with fixed topology (hence no $T_1$ transitions) have suggested, however,  that a possible underlying origin of this transition is the geometric incompatibility of the target  shape parameter with the embedding space: {in a regular version of the VM, rigidification occurs when the target shape parameter violates the isoperimetric inequality which requires $s_0\geq s_0^*(n)$, with $s_0^*(n)=\sqrt{4 \, n \tan(\pi/n)}$ for a regular $n$-sided polygon~\cite{farhadifar2007influence,staple2010mechanics,moshe2018geometric, merkel2018geometrically}.} An  ordered vertex model of $n$-sided polygons hence undergoes a transition  at $s_0=s_0^*(n)$. For $s_0<s_0^*$ the cells cannot achieve their target shape and the tissue is in a rigid, incompatible state,  with a single finite-energy ground state. For $s_0>s_0^*$ the tissue is soft/floppy, or compatible, with multiple zero-energy configurations.

In the incompatible and critically compatible state the tissue has a well-defined ground-state configuration, hence one may expect that such a ground state would be  a legitimate reference for measuring deformations and that an expansion about such a state to quadratic order in the strain would provide an accurate description of the linear elastic response of the system. 
In the present paper we show that this is not the case \mcm{at the critical point}, where the response of ordered VMs to small deformations deviates qualitatively from linear elasticity. 
In the next section we study the VM ground states and calculate the elastic moduli that quantify the response to uniform imposed loads. We then we focus on the critically compatible state where deviations from linear elasticity are most pronounced.
 
\section{Mean field theory and ground states}
\label{sec:MF}
The elastic moduli of a tissue encode information about the mechanical response to uniform external loads. We assume that in a uniform ordered tissue the responses of all cells are identical, and formulate a mean-field theory by considering the elastic energy of a single cell, Eq.~\eqref{eq:Ea}. To begin, we  express the energy in terms of configurational variables by introducing the symmetric $2\times 2$ metric tensor $\mathbf{g}$. Denoting the unit vectors defining a regular polygon by $\mathbf{v}_i$ and the polygon's area by $\Delta S$, we can then write
cell perimeter and area as
 \begin{eqnarray}
	A(\g) &=& \sqrt{\text{det}\,\mathbf{g}}~ \Delta S
	\;,
	\label{eq:area}\\
	P(\g) &=& 
	\sum_{i\in \mathrm{cell}} \sqrt{\rm{v}^{\alpha}_i g_{\alpha\beta}\rm{v}_i^{\beta}} \;,
	\label{eq:perimeter}
\end{eqnarray}
where Greek indices $\alpha,\beta$ denote Cartesian components. {Note that for triangles all configurations can be parametrized exactly as in Eqs.~\eqref{eq:area} and \eqref{eq:perimeter}. For higher order polygons the description of }
edges  in terms of a single uniform metric  is an approximation. 

 It is convenient to introduce dimensionless quantities by using $\sqrt{A_0}$ as the unit of length. The  dimensionless form of Eq.\eqref{eq:Ea} is then 
 \begin{equation}
 	E=\frac{E_{\text{cell}}}{\kappa_A A_0^2}=
 	\frac12\left[a(\g)-1\right]^2+ \frac{r}{2}\left[p(\g)-s_0\right]^2\;,
 	\label{eq:Energy}
 \end{equation}
with $a=A/A_0$, $p=P/\sqrt{A_0}$  and $r= \kappa_P/(\kappa_A A_0)$ a parameter that sets the relative cost of perimeter to area variations. 
This form of the energy functional has strong similarity with that of non-Euclidean shell theory, where stretching and bending terms may be incompatible due to violation of geometric compatibility conditions~\cite{siefert2021euclidean}. The absence of a stress-free configuration when $s_0<s_0^*$ is transparent in this form, since the isoperimetric inequality states that for $s_0<s_0^*$ no $\g$ can satisfy both $a=1$ and $p=s_0$ simultaneously.
 
\begin{figure*}
	\centering
	\includegraphics[width=\linewidth]{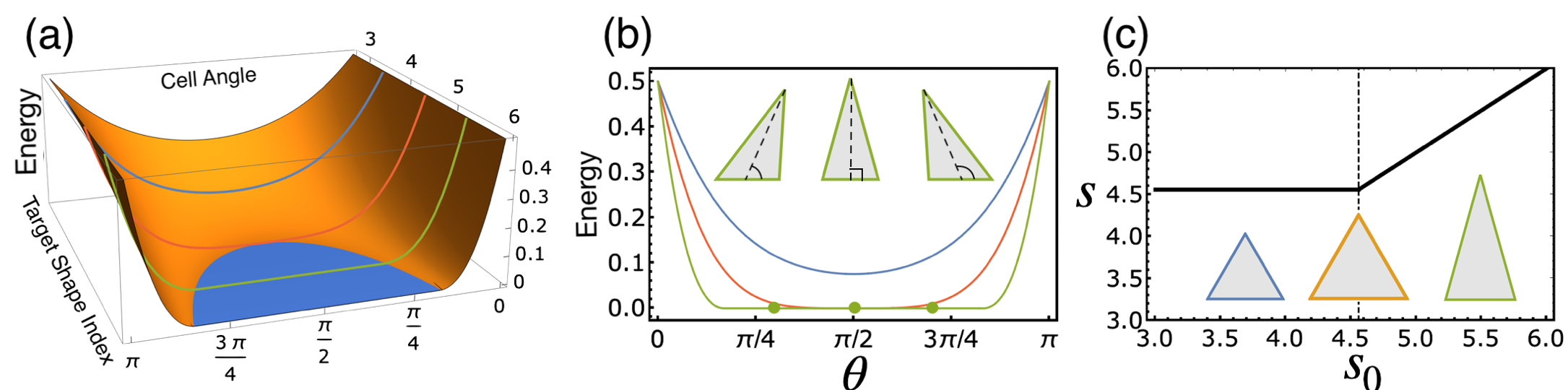}
	\caption{ (a) Ground-state energy $E_0$ as function of the target shape index $s_0$ and tilt angle $\theta$ measured between the median and the cell base. The blue region corresponds to zero energy, reflecting the degeneracy of the ground-state in the compatible regime. The colored lines describe $E_0$ as function of  $\theta$ for {$s_0=3.5,4.55901,5.3 $, corresponding to the incompatible, critical, and compatible regimes, respectively.} The three curves are plotted together in (b) as function of $\theta$, with the inset showing three different zero energy configurations corresponding to the compatible green curve. (c) {Observed {($s$)} vs. target perimeter {($s_0$)}.} In the incompatible regime $s_0<s_0^*$ the ground-state configuration is isotropic, whereas in the compatible regime $s_0>s_0^*$ the ground-state configuration is anisotropic and degenerate. The smaller scale of the incompatible (blue edge) cell illustrates the compromise of ground-state area and perimeter being smaller/larger relative to their target values $a<1$, $p>s_0$.}
	\label{fig:degenerateshapes}
\end{figure*} 
 
 
Before examining the mechanical response to small perturbations, we need to find the  ground-state energy. This is determined by   minimizing Eq.~\eqref{eq:Energy} with respect to all admissible metric tensors $\mathbf{g}$. The ground state metric $\go$ is given by
 \begin{equation}       
	\go = \arg\min_{\g} E(\g ;s_0,r) \;.
	\label{eq:grdstateminimization}
\end{equation}

The calculation of  $\go$  for a given $n$-sided regular polygon as a function of $(s_0,r)$ can be carried out analytically and is shown in \appref{app:GS} for $n=3,4,6$. For $s_0 \leq s_0^*$ there is a unique ground state corresponding to a regular polygon (and a gapped energy if $s_0 < s_0^*$). In this regime, referred to as the incompatible regime, the system is rigid. As $s_0\rightarrow s_0^*$ from below, the energy gap vanishes. For $s_0>s_0^*$ the system transitions to the compatible regime, where there is a one-parameter set of zero-energy configurations, making the tissue floppy.
This is shown in \figref{fig:degenerateshapes}(a,b) where we plot the ground-state energy of a single cell as a function of its target shape parameter $s_0$ and the tilt angle between the median and the cell base, which provides a measure of shear deformation. This angle parametrizes a family of zero energy states in the compatible regime, \mcm{as shown explicitly in Appendix~\ref{app:GS}.}  In \figref{fig:degenerateshapes}(c) we show the observed ground state shape parameter $s$ as a function of the target shape parameter $s_0$. The inset displays ground-state configurations. In the incompatible regime for $s_0<s_0^*$,  $s=s_0^*$. In the compatible regime the system can achieve both target area and perimeter, with a family of tilted polygonal shapes satisfying  $s=s_0$, corresponding to the flat region in \figref{fig:degenerateshapes}(a,b). The smaller scale of the incompatible cell in \figref{fig:degenerateshapes}(c) reflects the compromise between area and perimeter costs resulting from $a<1$ and $s>s_0$.
\section{Linear Response to Mechanical Deformations}
\label{sec:response}
In this section we examine the response of the VM to small mechanical deformations.
\mcm{It is useful to first}  consider a conventional elastic solid described by an energy $E(\g)$ with a unique ground state $g_{0\mu\nu}=\delta_{\mu\nu}$ that provides the reference (undeformed) configuration. The  mechanical response to a deformation is quantified in terms of the strain $\mathbf{u}$ defined by writing $\g = \go + 2 \mathbf{u}$. Linear elasticity can then be formulated by expanding the energy around the reference state to quadratic order in the strain as
 \begin{equation}
 		E(\g)=E(\go+2\mathbf{u})=
 		 \frac{1}{2} A^{\alpha\beta\gamma\delta}(\go) u_{\alpha\beta} u_{\gamma\delta} + O(\mathbf{u})^3\;,
	\label{eq:LinearEnergy1}
\end{equation}
with $E(\go)=0$. For an isotropic solid, as well as for a triangular lattice, the elastic stiffness tensor $A^{\alpha\beta\gamma\delta}$ has the form
\begin{equation}
	A^{\alpha\beta\gamma\delta} = \lambda  \go^{\alpha\beta} \go^{\gamma\delta} +  \mu \left(  \go^{\alpha\gamma} \go^{\beta\delta} + \go^{\alpha\delta} \go^{\beta\gamma}\right)\;,
	\label{eq:ET}
\end{equation}
and \mcm{is fully specified in terms of two} independent \mcm{quantities}, the Lam\'e coefficients $\lambda$ and $\mu$. The elastic moduli characterizing the linear response to any deformation can then be expressed in terms of  $\lambda$ and $\mu$, according to the expressions given in the last column of Table 1.


\subsection{Incompatible regime}
We begin by analyzing the mechanical response in the incompatible regime where linear elasticity holds. 
To \mcm{evaluate the elastic constant of the  VM in the incompatible regime} we first identify the
unique  ground state configuration with respect to which deformations are measured $g_{0\mu\nu} = c^2 \delta_{\mu\nu}$, which corresponds to a regular $n$-sided polygon. The constant $c$ is determined by energy minimization and is the real solution of a cubic equation, given in  \appref{app:GS} for $n=3,4,6$. 
We then expand \eqref{eq:Energy} in powers of $ \mathbf{u} = \tfrac{1}{2}(\g-\go)$ as in \eqref{eq:LinearEnergy1}. Since $\go$ is isotropic, the elastic stiffness tensor $A^{\alpha\beta\gamma\delta}$ has  the form given in Eq.~\eqref{eq:ET} and is entirely determined by the  two coefficients $\lambda$ and $\mu$. \mcm{For a triangular polygon these are} given by 
\begin{equation}
 	\begin{split}
 		\lambda  &= \frac{\sqrt{3}}{2} + \frac{9 r}{4 c}  (3 s_0 - 7c ) \;,\\
 		\mu  &= \frac{9 r}{4 c}  (3 c -s_0 )\;.
 		\label{eq:LameVM}
 	\end{split}
\end{equation}

\begin{figure}
	\centering
	\includegraphics[width=0.8\linewidth]{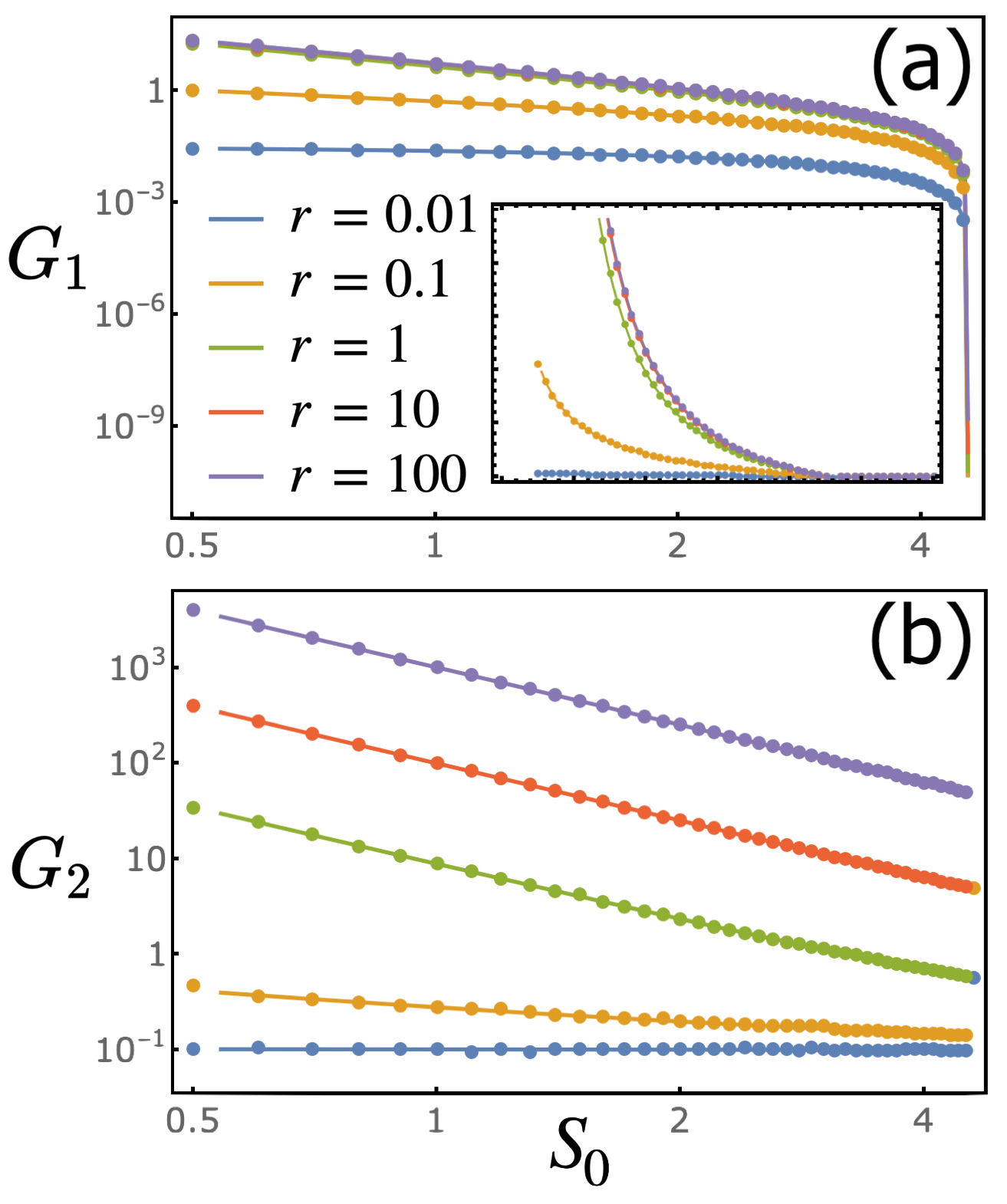}
	\caption{Analytical (solid) and numerical (points) elastic moduli $G_{1,2}$ for uniaxial deformation with transverse direction free (a) or clamped (b),  of an $n=3$ VM as functions of target shape parameter $s_0$ on a log-log scale. {The compatible/incompatible transition is at $s_0^*=2\ 3^{3/4}$.}  The inset of (a) shows $G_1$ on a linear scale,  highlighting the vanishing of the measured response beyond the critical shape parameter.
	The moduli were calculated for various values of rigidity ratio, ranging from $r = 0.01$ to $r=100$.}
	\label{fig:Moduli}
\end{figure}
\mcm{The corresponding expressions for the Lam\`e coefficients} for hexagonal tissue are given in \mcm{Eq.~\eqref{eq:lame_hex}.}  The elastic moduli describing specific deformation protocols \mcm{can then be obtained using the relations given in} Table~\ref{table:1}. 
The elastic constants calculated analytically agree well with the results of numerical simulations, 
as shown in \figref{fig:Moduli}. 

\subsection{Compatible regime}
\mcm{In the compatible regime linear elasticity  fails because the ground state is degenerate, as} shown  in \figref{fig:degenerateshapes}(a), where the flat region corresponds to a continuous set of rest configurations \cite{moshe2018geometric,kupferman2020continuum}. This means that when subject, for instance, to a small uniaxial deformation, the system can accommodate the deformation by \mcm{changing its shape and }finding a new zero energy configuration corresponding to the deformed shape, resulting in vanishing elastic constant $G$. The elastic constants corresponding to a specific deformation can still be calculated using the procedure defined in  Eq.~\eqref{eq:LinearEnergy1} and vanish whenever the  deformed state corresponds to one of the degenerate ground states.
\mcm{This procedure, however, fails at the boundary of the manifold of degenerate ground states shown in blue in Fig.~\ref{fig:degenerateshapes}a.}
On this boundary, the elastic constants cannot be calculated using Eq.~\eqref{eq:LinearEnergy1}  since they are sensitive to the sign of deformation, with vanishing constants for deformations \mcm{that displace the system}  towards the blue region, and finite constants \mcm{for deformations that displace it in} the other direction.



In the next section we examine the response at the critical \mcm{point separating} the compatible and incompatible states. \mcm{We show that at the onset of rigidity
the VM exhibits anomalous elasticity, which arises directly from 
the nonanalyticity of the energy functional.}
\begin{table}[h!]
\centering
\begin{tabular}{ |c|c|c|c|c|  }
 \hline
 \multicolumn{5}{|c|}{Tissue Moduli} \\
 \hline
  & Deformation & Fixed &  Free & Linear elastic solid\\
 \hline
 $G_1$  & Uniaxial & $u_{11}$ & $u_{12},u_{22}$ & $ \tfrac{4 \mu (\lambda + \mu)}{\lambda+2\mu}$ \\
 $G_2$ & Uniaxial & $u_{11},u_{22}=0$ & $u_{12}$& $\lambda + 2\mu$\\
 $G_3$ & Area  & $u_{11} = u_{22}$    & $u_{12}$& $4 (\lambda + \mu$)\\
 $G_4$&  Shear & $u_{11} = -u_{22}$    & $u_{12}$ & $4 \mu$\\
 $G_5$ & Shear &$u_{12}$ & $u_{11},u_{22}$& $4 \mu$\\
\hline
\end{tabular}
 \caption{Elastic moduli for five different deformation protocols. The last column shows the expressions in terms of the Lam\'e coefficients $\lambda$ and $\mu$  for the case of  a linear elastic solid in $2D$, where $\mu$ is the shear modulus and $\lambda+2\mu$ the compression modulus.}
\label{table:1}
\end{table}

\section{Breakdown of linear elasticity at the critical point}
\label{sec:Violation}
   

We now \mcm{examine the mechanical response of the VM at the critical point corresponding to $s_0=s_0^*$.
We focus specifically on the triangular VM, but the same behavior occurs generically for all polygonal shapes.}
\mcm{At the critical point there is a single compatible ground state configuration with} $a=1$ and $p=s_0$, corresponding to an equilateral triangle. {The associated ground state has  zero energy and is unique.} 

\mcm{Since the energy is nonanalytic at the critical point, the elastic constants cannot be evaluated by expanding the energy for small deformations. Instead we calculate them by examine the response to the various deformation protocols summarized in Table 1. The dependence of the elastic constant on the specific protocol demonstrates the nonanalyticity of the energy functional and the failure of linear elasticity.}

\begin{figure*}
	\centering
	\includegraphics[width=\linewidth]{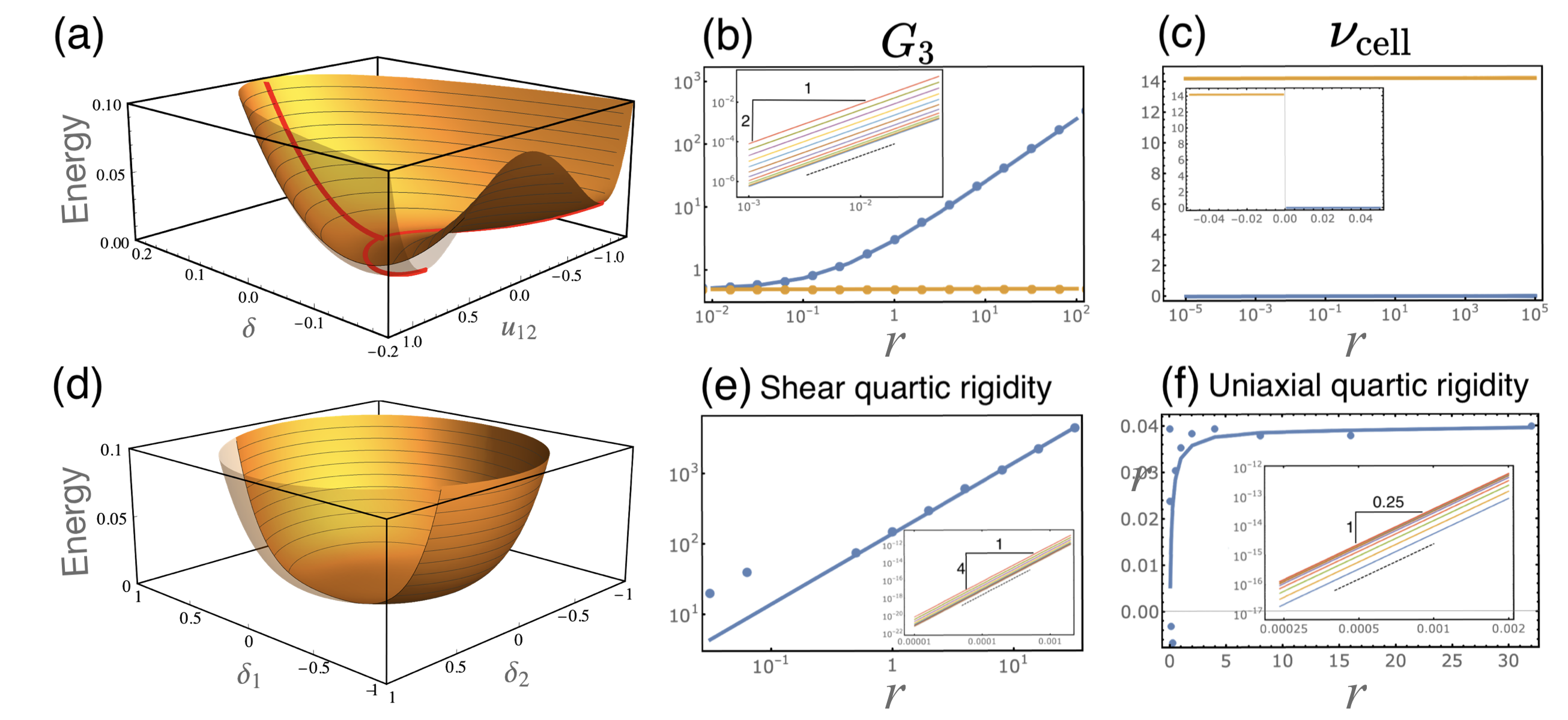}	\caption{Energy landscape and mechanical properties of critically compatible triangular cell model. (a) Energy as function of imposed area and shear strains, showing an asymmetric response to area compression and tension. (b) Resistance to area deformation as function of rigidity ratio for tensile (blue) and compressive (yellow) area strain, compared with VM numeric results (dots). Inset shows the log-log plots of energy-strain curves. (c) The cell-ratio defined in \eqref{eq:cellratio} as function of rigidity ratio. Inset shows the cell ratio as function of imposed strain for a given rigidity ratio, confirming that it is well defined material property. (e) Energy landscape for area preserving deformations as function of the two shear strain modes, presenting a flat landscape with vansishing quadratic rigidity and finite quartic rigidity. (e) Quartic rigidity as function of rigidity ratio, with inset showing log-log plots of the energy-strain curves confirming the quartic dependence of energy in strains. (f) Quartic order rigidity as function of rigidity ratio (quadratic rigidity vanishes) compared with VM numeric results (dots). Inset shows the energy-strain curves on a log-log scale validating the quartic dependence of energy in strain.   }
	\label{fig:Crit}
\end{figure*}

\mcm{To demonstrate this, we begin by showing that}   the response to area deformations as measured by the modulus $G_3$ in Table \ref{table:1} \mcm{is asymmetric, in the sense that the response to isotropic compression is different from the response to isotropic extension. To calculate $G_3$ we}  impose $u_{11} = u_{22} = \delta$ and allow $u_{12}$ to be selected by energy minimization.  
The plot of the energy as a function of imposed area strain $\delta$ and spontaneous shear strain $u_{12}$ shown in
\figref{fig:Crit}(a) \mcm{reveals the origin of the asymmetry.}
The red curve represents the energy minimizer for a fixed area deformation $\delta$. \mcm{It is evident that tensile deformations, corresponding to $\delta>0$, maintain $u_{12}=0$, hence} induce no  shear, while compression, corresponding to $\delta <0$, yields a finite value of $u_{12}$, hence induce spontaneous  shear. The bifurcation of the red curve at the global minimum {indicates spontaneous symmetry breaking in the shear response. 
The \mcm{plot of $G_3$} as a function of the rigidity ratio $r$ }  for compressive (yellow) and tensile (blue) deformations in \figref{fig:Crit}(b) clearly shows the asymmetry. The inset shows log-log plots of energy-{strain} curves for the tensile case, demonstrating  the quadratic dependence of energy on strain. These findings confirm the non-analyticity of the energy functional at the \mcm{critical point} ground-state.

To quantify the magnitude of the \mcm{spontaneous shear induced}  in response to compressive area deformations, we define the cell ratio $ \nu_\text{cell}$ in analogy to the Poisson ratio as 
\begin{equation}
    \nu_\text{cell} =u_{12}^2/\delta\;.
    \label{eq:cellratio}
\end{equation}
This definition is chosen instead of the na\"ive measure $u_{12}/\delta$ because the latter is found to depend on the magnitude $\delta$  of the imposed strain and thus it is not a well defined material property. 
The \mcm{cell ratio is shown} in \figref{fig:Crit}(c) \mcm{as a function of the stiffness ratio $r$. It clearly} captures the asymmetry between tensile and compressive deformations,  with $\nu_\text{cell}=0$ for tensile forces and $\nu_\text{cell}\not=0$  for compressive forces. 
The inset of \figref{fig:Crit}(c) shows $\nu_\text{cell}$ for fixed $r$ as a function of $\delta$, confirming that this parameter is indeed a well defined material property independent of  $\delta$.
The cell-ratio quantifies the coupling between bulk and shear deformations, which is absent in an isotropic linear elastic solid. 

Next, we study the response to (area preserving) pure shear deformations by imposing $u_{12}=\delta$ and letting $u_{11}=-u_{22}$ to be selected by energy minimization, or vice versa. The plot of the energy as a function of shear strain shown in \figref{fig:Crit}(d) shows that {the two shear modes}  are decoupled as in classical elasticity. A log-log plot of the energy-strain curve for the trace-less shear mode $u_{11}=-u_{22}$ and various values of $r$ shown in inset of \figref{fig:Crit}(e) reveals an inherently nonlinear quartic dependence on strain, demonstrating on the importance of nonlinear effects for infinitesimally small loads.
The quartic rigidity is plotted in \figref{fig:Crit}(e) on a linear scale. The disagreement between theory and simulations at low rigidity ratio reflects a failure of convergence of the energy minimizing gradient-descent algorithm.

Finally, we evaluate the response to a uniaxial strain and discover that, similar to the shear response, the quadratic rigidity vanishes and the response is quartic. The quartic rigidity is plotted as function of rigidity ratio in \figref{fig:Crit}(f) and the inset shows the energy-deformation curves on log-log scale.

In summary, we have shown in this section that  in {the critically} compatible state linear elasticity fails to describe the linear response of the VM to small deformations. First, the asymmetric response to tensile and compressive loads  
violates reciprocity.
Second, the response to shear deformations reveals quartic rigidity, violating the superposition principle {even for infinitesimal deformations}. 
Finally, we uncovered an anomalous coupling between area and shear deformations, with a spontaneous breaking of symmetry in the shear response to isotropic dilations. This is reminiscent of the recently discovered odd-ratio  that quantifies area-shear coupling in a generalized linear elasticity of active solids \cite{scheibner2020odd}.

These findings are also related with the recently suggested framework of energetic rigidity \cite{damavandi2021energeticI,damavandi2021energeticII}. Within this framework, a system is termed (energetically) rigid if a finite deformation increases energy at any order, not necessarily quadratic one. According to this definition the quartic shear-rigidity is a signature of energetic rigidity. The emergence of sign-dependent response, violation of reciprocity and bulk-shear coupling indicate the non-analyticity of the energy functional and therefore \mcm{shows that one cannot describe VM elasticity through a Taylor expansion of the energy for small deformations.}  Specifically, we have shown that at the critical point the VM can introduce different rigidities for tensile and compressive loads. 
 In the next section we explore the origin of the failure of linear elasticity using a visual representation of the problem. 

\section{Visual representation of the failure of linear elasticity}
\label{sec:Visual} 
 \subsection{Elastic triangle}
 \begin{figure*}
	\centering
	\includegraphics[width=0.9\linewidth]{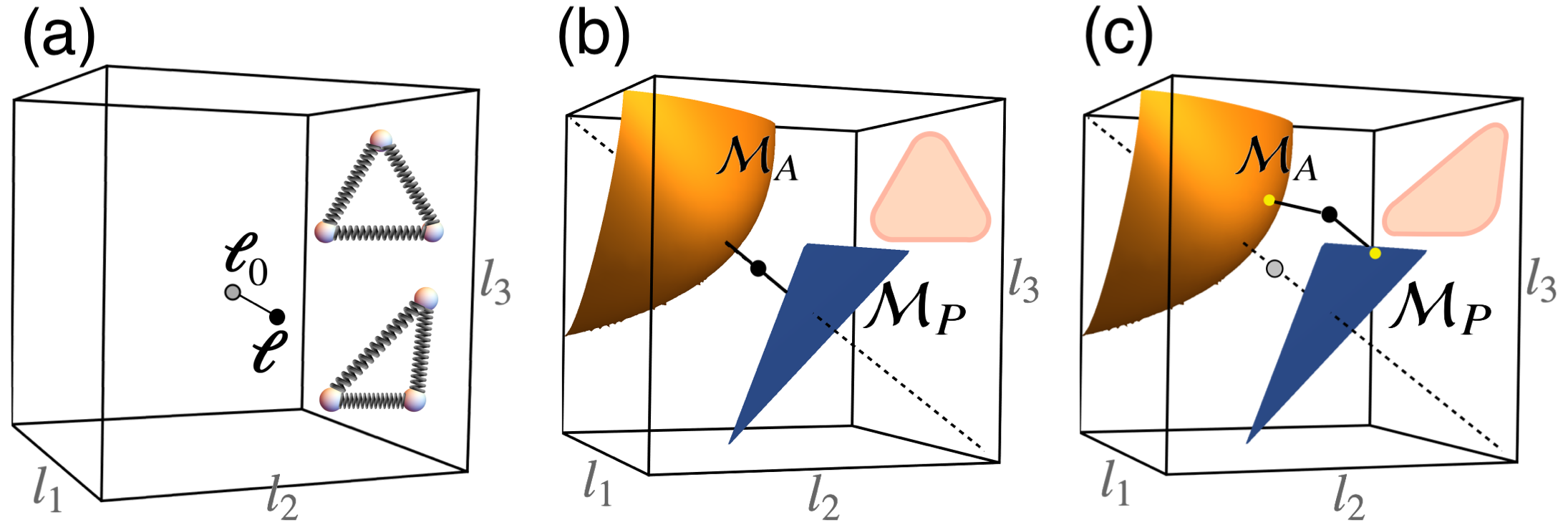}
	\includegraphics[width=0.9\linewidth]{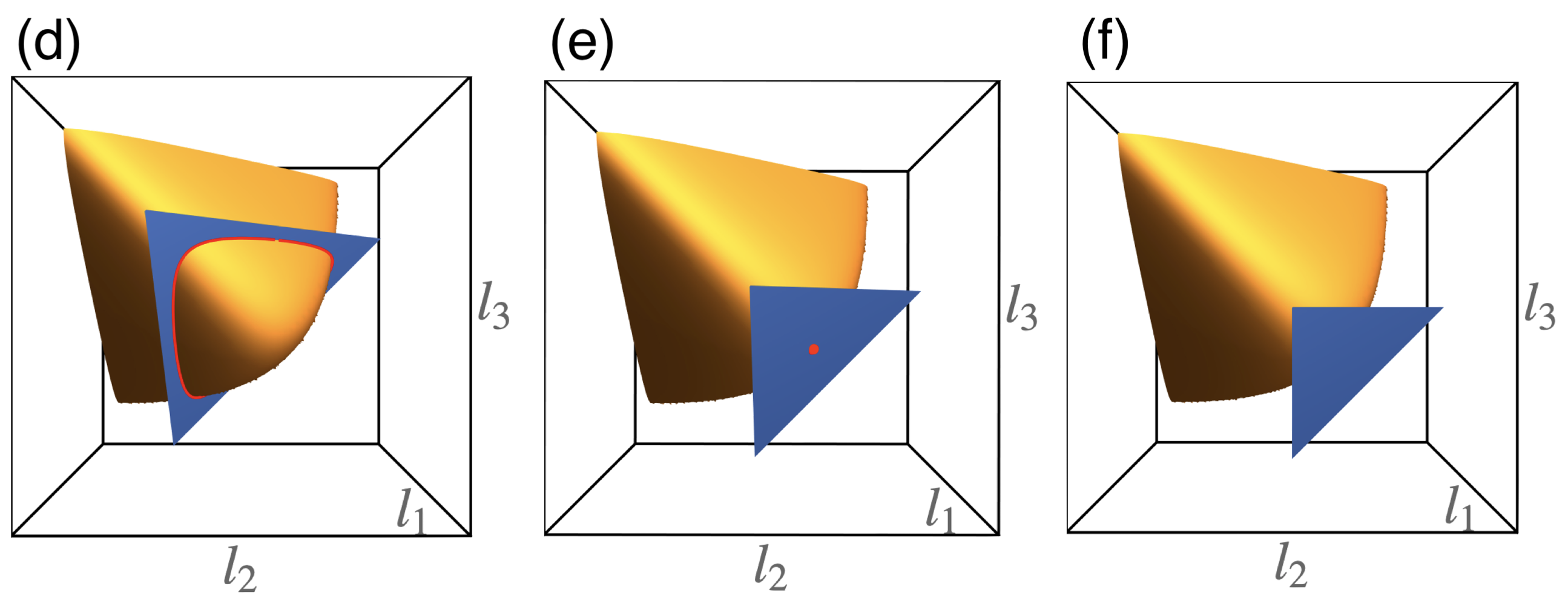}
	\caption{Visualizing elastic and cell models in configuration space. (a) An elastic triangle made of masses and springs. The rest and deformed configurations are marked by the gray and black points correspondingly. 
	The mechanical energy is a measure for the distance of the configuration $\boldsymbol{\ell}$ from $\boldsymbol{\ell}_0$.
	(b) A triangular cell model with rest area and perimeter. $\Mp$ is the set of all configurations satisfying the rest perimeter and $\Ma$ is the set of all configurations satisfying the rest area. The black point is the energy ground-state with its exact position depending on the rigidity ratio. (c) a deformed triangular cell model, illustrating the content of the mechanical energy is measuring the joint distance from the two surfaces. The yellow points are hidden internal degrees of freedom that are selected to minimize the distance from $\Mp$ and $\Ma$.
		(d) Floppy cell with $s_0^* <s_0$: The intersecting curve represents a continuous set of triangles satisfying both $P_0$ and $A_0$ simultaneously. (e) Critically rigid cell with $s_0^*=s_0$ having one configuration that satisfy are and perimeter simultaneously. (f) $\Ma$ and $\Mp$ are disjoint, hence no triangle can satisfy both conditions and it is therefore frustrated.}
	\label{fig:ToyIllustration}
\end{figure*}
{To introduce a pictorial representation of deformations in configuration space we first consider a common microscopic model for elastic solids,} which is a lattice of masses and springs. In $2D$ a triangular lattice of identical masses and springs leads, in the coarse-grained limit, to homogeneous and isotropic linear elasticity~\cite{seung1988defects, kupferman2018variational}. As discussed before, the response to uniform loads is equivalent to the response of a single triangle. Therefore we consider a single triangle made of three identical masses and harmonic springs with rest lengths $l_0$. The rest configuration forms a point in configuration space, denoted ${\bm\ell_0} = \left(l_0,l_0,l_0\right)$ and shown as a gray point and associated equilateral triangle in \figref{fig:ToyIllustration}(a).
An arbitrary deformed state is denoted by ${\bm\ell} = \left(\ell_1,\ell_2,\ell_3\right)$ and shown as a black point and associated deformed triangle in \figref{fig:ToyIllustration}(a). Deformation along the $\mathbf{n}_a = \tfrac{1}{\sqrt{3}}(1,1,1)$ direction correspond to area deformation, i.e.,  response to pressure changes, and deformations along the perpendicular plane spanned by $\mathbf{n}_1 = \tfrac{1}{\sqrt{2}}(1,-1,0), \mathbf{n}_2 = \tfrac{1}{\sqrt{2}}(0,1,-1)$ correspond to shear deformations.
Deviations from the rest configuration cost energy proportional to $\delta {\bm\ell}^2$, {with $\delta\bm\ell=\bm\ell-\bm\ell_0$.} When expressed geometrically the rest and actual configurations can be represented by reference and actual metrics $\g_0$ and $\g$, respectively, and the energy can be expanded in powers of $u = \tfrac{1}{2}(\g - \g_0)$ as in Eq.~\eqref{eq:LinearEnergy1},
with $A^{\alpha\beta\gamma\delta}$ as in Eq.~\eqref{eq:ET}.
Importantly, the energetic response to a generic deformation along a given direction in configuration space is insensitive to the orientation, as expected from a quadratic expansion. In addition, there is no coupling between bulk and shear deformations; for example $A^{1112} = 0$.

\subsection{VM triangle}
We now implement the same visual representation described above for an elastic triangle for the case of a triangular VM cell, that is a triangle defined by its target area and perimeter. Contrary to the elastic triangle, the terms in the cell energy Eq.~\eqref{eq:Ea} penalize  geometric deformations of area and perimeter, which {do not uniquely determine a configuration of a triangular polygon.}
The area term penalizes deviations from the target area, {which identifies a $2D$ manifold of equal area configurations denoted by $\Ma$ and shown as an} orange surface in \figref{fig:ToyIllustration}(b,c). The perimeter term penalizes  deviations from the target perimeter, which identifies a $2D$ manifold of equi-perimetric configurations $\Mp$ shown as a blue surface in \figref{fig:ToyIllustration}(b,c).
The black point in \figref{fig:ToyIllustration}(b) represents the ground-state configuration that is achieved in the incompatible regime by {balancing} area and perimeter deviations. Contrary to classical elasticity which measures the distance of a point in configuration space from a reference point, the cell energy measures the joint distance from two target surfaces. This introduces additional \MM{hidden} degrees of freedom to the deformations,  as shown in \figref{fig:ToyIllustration}(c) where the energy of the deformed configuration (black point) is measured by selecting the closest (yellow) points on the target manifolds $\Ma, \Mp$. 

The state of the tissue is determined by the relative location of the two surfaces in configuration space. In \figref{fig:ToyIllustration}(d-f) we show three different situations where the two surfaces cross each other along a curve, at a point, or not at all, corresponding to floppy ($p_0^*<p_0$), critically rigid ($p_0^*=p_0$), and frustrated cell ($p_0<p_0^*$).
The  ground-state is a point located along the $\mathbf{n}_a$ direction in between the surfaces, with its exact position depending on the rigidity ratio $r$: for $r\gg 1$ ($r\ll 1$) the cell is dominated by perimeter (area) term and the ground-state is closer to $\Mp$ ($\Ma$).
Zero energy states exist only if the two surfaces intersect as in \figref{fig:ToyIllustration}(d,e). When the two surfaces are disjoint as in \figref{fig:ToyIllustration}(f), the joint distance of any point in space from the surfaces, that is the energy, is necessarily non-zero, reflecting the energy gap and the emergent rigidity of the cell. 

\begin{figure}
	\centering
	\includegraphics[width=\linewidth]{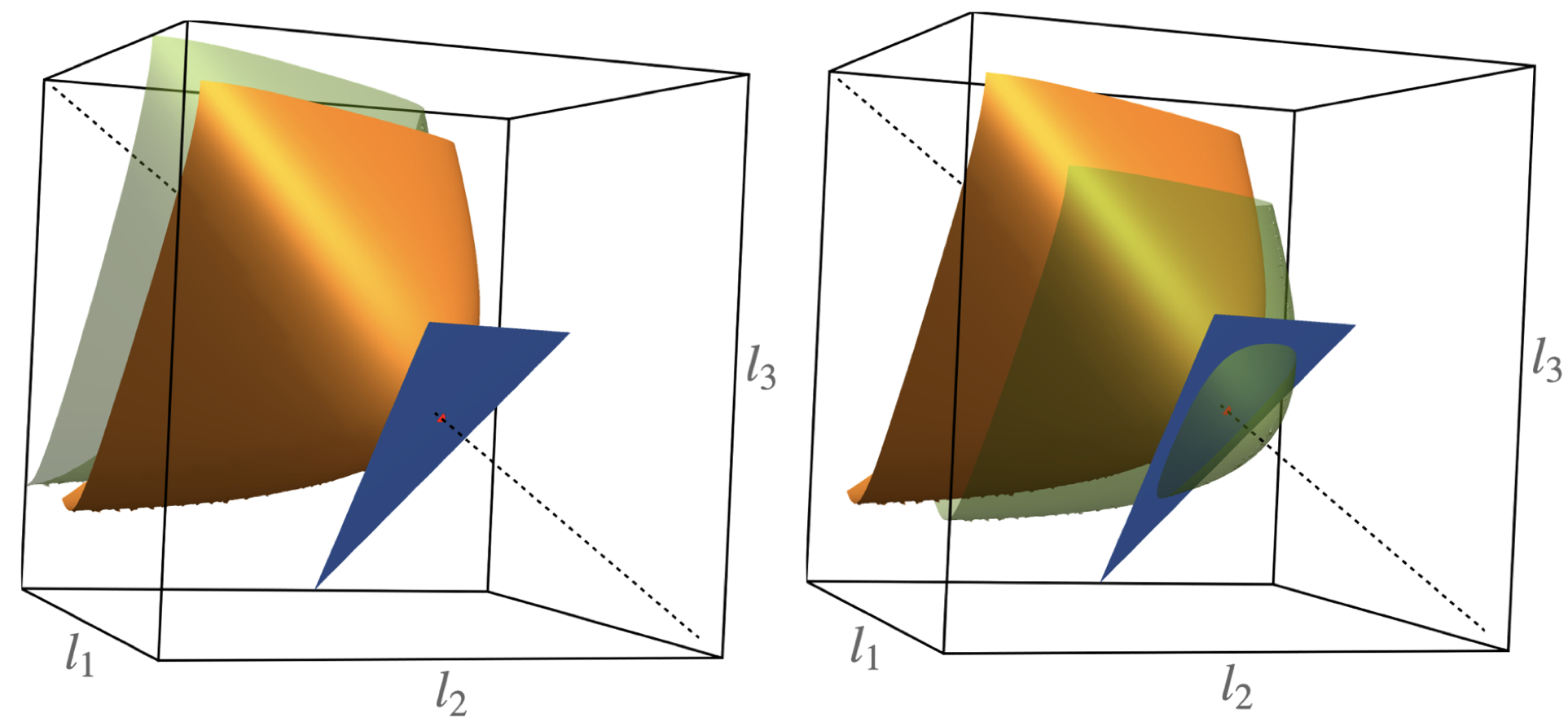}
	\caption{Area deformation in a critically rigid triangular cell. The target surfaces $\Ma$ and $\Mp$ intersect at a point marked by red dot, forming the zero energy ground-state. Area deformation is imposed by constraining the configuration to a different equi-areal surface. (a) Area tension - the equilibrium is positioned along the (1,1,1) direction, between the constraining green surface and the target perimeter surface $\Mp$, hence isotropic, with both area and perimeter energy deviations. (b) Area compression - the constraining green surface intersects with the target perimeter $\Mp$. The selected configuration is therefore on the intersecting curve with zero perimeter energy and induced shear strain. }
	\label{fig:ShearInstabilityIllustration}
\end{figure}

It is then evident why a critically compatible tissue present anomalous elasticity.
Assume a critically compatible triangle with target perimeter $P_0 = 3$ and target area $A_0 = \sqrt{3}/4$. 
The triangle in this case is compatible, and there is only one configuration satisfying $A_0$ and $P_0$ simultaneously: an equilateral triangle of edge length $l=1$, with zero energy. 
In \figref{fig:ShearInstabilityIllustration}(a,b) the orange and blue surfaces represent the target area and perimeter surfaces, and intersect  at the single point corresponding to the ground-state. Now consider an infinitesimal area deformation. Area expansion corresponds to constraining the cell configuration to lie on the green surface in \figref{fig:ShearInstabilityIllustration}(a). In this case the perimeter necessarily deviates from its target value, and the closest point on $\Mp$ remains isotropic. {In contrast, an area compression corresponds to the situation shown  in \figref{fig:ShearInstabilityIllustration}(b), where the green surface describing the deformed area and the target perimeter surface intersect,} resulting in zero perimeter energy and {finite degenerate} area energy. {Therefore the system spontaneously breaks the symmetry by selecting  a deformed state  corresponding to finite  shear of fixed magnitude and arbitrary orientation.} Also, the resistance to area compression depends only on area rigidity whereas area tension depends on both area and perimeter rigidities. This is in complete agreement with the analytical and numerical results obtained in \figref{fig:Crit}.

Finally, the visual representation in \figref{fig:ShearInstabilityIllustration}  {clarifies why the definition} of the cell-ratio given in Eq.~\eqref{eq:cellratio} is independent of the imposed strain and {constitutes} a material property. 
\figref{fig:ShearInstabilityIllustration}(b) shows that the imposed area strain $\delta$ measures the translation of the green surface, and the induced shear strain is the distance between the undeformed state, marked by the red point, {and the  curve where the green surface and the blue manifold $\Mp$ intersect.} For small $\delta$ the part of the green surface that intersect with $\Mp$ can be approximated as a spherical cup. The relation between its radius of curvature $R$, and the imposed and induced strains is 
\begin{equation}
    (R-\delta)^2 + u_{12}^2 = R^2
\end{equation}
and for small $\delta$ we get
\begin{equation}
 \nu_\text{cell}= \frac{u_{12}^2}{\delta} = 2 R  \ .
\end{equation}
{The cell-ratio quantifying the coupling between imposed area strain and induced shear strain is thus} a geometric measure of the curvature of $\Ma$. The numeric simulations and analysis confirm that this definition is well defined and independent of the imposed strain magnitude, as shown in \figref{fig:Crit}(c).


\section{Summary and Discussion}
\label{sec:summary}

In summary, we have shown that while in the incompatible regime the VM obeys linear elasticity, \mcm{qualitative deviations from linear elasticity} are found at the onset of mechanical rigidity for $s_0=s_0^*$, \mcm{including nonreciprocal response to isotropic area changes and spontaneous shear upon isotropic dilation. These deviations are unexpected, given the critical state has a single non-degenerate ground state, and demonstrate the non-analytic nature of the energy functional at the critical point. The compatible fluid-like regime for 
 $s_0>s_0^*$  also exhibits non-Hookean elasticity, but this is due to the existence of  a continuum of degenerate ground states that allow the system to accommodate external deformations at no energetic cost by changing its shape.}


To understand the mechanisms that drives the failure of linear elasticity in the critically compatible case, we have developed a graphic representation of the mean-field of the VM that  illustrates the existence of hidden degrees of freedom.
This geometric representation shows that the elastic solid holds two distinct sets of reference configurations (associated with target area and target perimeter) that may be either compatible or incompatible with each other. 
 When compatible, the system is fluid in the sense that it can explore a manifold of degenerate zero energy states and accommodate deformations with no energetic cost, below a critical strain. When the two reference states are incompatible, the system is rigid and has a finite ground state energy determined by the distance between the two sets of reference states that cannot be simultaneously accommodated. The existence of this finite energy or pre-stress provides a definition of geometric rigidity. 
The deviations from linear elasticity occur at the critically compatible state, where the system has a single non-frustrated ground-state, yet  reciprocity is violated, an anomalous coupling between bulk and shear deformations emerges, and quartic rigidity is observed in response to specific deformations.


In the present work, we have restricted ourselves to a mean-field theory that examines the linear response of the VM to spatially uniform deformations, where all cells respond in the same way. The identification of hidden degrees of freedom demonstrates that analyzing the response of nonlinear and non-uniformly deformed tissue, e.g., the response  of the tissue to the \mcm{localized} contraction of a single cell, requires  a generalized elastic  framework.

The relevance of our work goes beyond the scope of tissue mechanics in two main directions.
First, our work provides a new route for the design of mechanical meta-materials with extreme properties. Specifically, the unusual mechanical properties of the tissue VM stem directly from the geometry of the reference surfaces in \figref{fig:ToyIllustration}. This suggests that one could design materials with extreme mechanical behavior by constructing  a cellular network where each cell has a specific local energetic response, controlled by  the geometry of the reference surfaces. 
Second, the well-established paradigm in physics that response to small perturbations can be  analyzed via a Taylor's expansion about the ground state, fails at criticality, as indicated by the assymetric response to tensile and compressive area deformations. 
Our work suggests a new framework for formulating  the elasticity of underconstrained system by describing them via analytic-like quadratic energy functionals where the available (and possibly incompatible)  reference states are incorporated as dynamical fields. The identification of  ground states and elastic response then requires additional minimization with respect to such reference states.
These results and observations provide independent support for the earlier model proposed in \cite{moshe2018geometric}.

\section{Acknowledgements}
{We} thank Max Bi for illuminating discussions {and for providing the original version of the code used in the simulations.} This research was supported in part by the National Science Foundation under Grant No. NSF PHY-1748958 (M.J.B.), Grant No.~DMR-2041459 (A.H. and M.C.M.), by the Israel Science Foundation grant No. 1441/19 (M.M.), and through the Materials Science and Engineering Center at UC Santa Barbara, DMR-1720256 (iSuperSeed) (M.C.M. and M.J.B.).

\appendix

\section{VM numerical simulations}
\label{app:FE}

To test the analytical results, we have  simulated numerically a VM with a regular lattice of triangular cells, implementing the model in Surface Evolver~\cite{brakke1992surface}. Periodic boundaries are used to avoid boundary effects, with periodic lengths $L_x$, $L_y$ and a shear length $L_{xy}$ such that $(x, y) = (x + m L_x + n L_{xy}, y + n L_y)$, where $m$ and $n$ are integers. For a given shape index $s_0$ and rigidity ratio $r$ we first find the ground state using a gradient descent method to minimize energy over the vertex positions and the periodic boundary lengths $L_x$ and $L_y$, with $L_{xy} = 0$.

\begin{figure}
    \centering
    \includegraphics[width=0.5\textwidth]{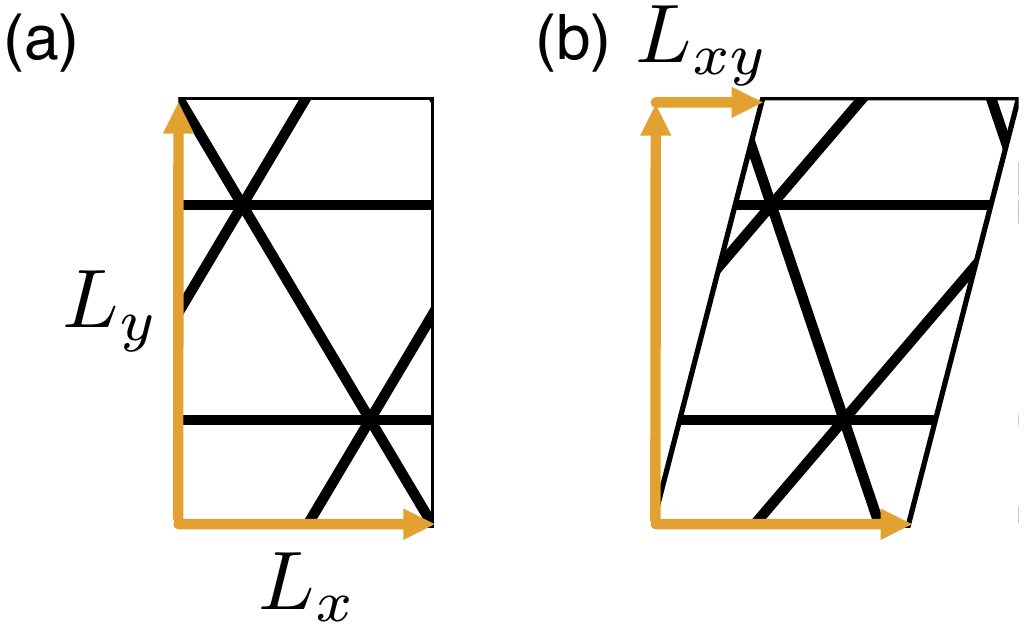}
    \caption{Numerical simulations. (a) Image of numerical simulations with no shear. The periodic boundary are defined by a parallelogram with horizontal length $L_x$ and vertical length $L_y$. (b) Image of a sheared tissue, with shear length $L_{xy}$}
    \label{fig:vm_sims}
\end{figure}

From this ground state, we calculate the tissue moduli $G_i$, $i = 1 ... 5$, using the same procedures for $u_{11}$, $u_{12}$, and $u_{12}$ (see Table~\ref{table:1}). The periodic lengths are transformed as $L_x \rightarrow L_x (1 + u_{11})$, $L_y \rightarrow L_y (1 + u_{22})$, and $L_{xy} \rightarrow u_{12}L_y$. We then minimize energy under this strain by updating vertex positions and free strain parameters. The modulus $G$ is then calculated as $E = E_0 + \frac{1}{2}G \delta^2$ where $E$ is the mean energy per cell, $E_0$ is the ground state energy per cell, and $\delta$ is the strain magnitude. Unless otherwise state, a value of $\delta = 0.001$ is used.
In \figref{fig:vm_sims} we show the energy minimizing configuration of a unit cell before and after shear deformation. 

\section{Analytical calculation of ground states }
\label{app:GS}
{It is instructive to display the calculation of the ground states for a quadrilateral ($n=4$), where the isoperimetric ratio is $s_0^*=4$. In this case the derivation is transparent and  can be carried out analytically.}

{The metric tensor can be parametrized in terms of the (dimensionless) lengths $c_1$ and $c_2$ of opposite parallel sides of the quadrilateral and the angle $\theta$ between adjacent sides as 
\begin{equation}
g_{ij}=
\begin{pmatrix}
c_1^2&c_1c_2\cos\theta\\
c_1c_2\cos\theta & c_2^2
\end{pmatrix}
\label{eq:metricmatrix}
\end{equation}
with $p=2(c_1+c_2)$ and $a=c_1c_2\sin\theta$. We choose  $0\leq\theta\leq\pi/2$. Inserting this into the mean-field  energy Eq.~\eqref{eq:Energy}, we obtain
\begin{equation}
E= \frac12\left(c_1c_2\sin{\theta}-1\right)^2+ \frac{r}{2}\left(2(c_1+c_2)-s_0\right)^2.
\label{eq:squareenergy}
\end{equation}
The ground states are obtained by finding the metric that minimizes the energy. This gives three equations in three unknowns, 
\begin{equation}
\begin{aligned}
\frac{\partial E }{\partial c_1}&= \left(c_1c_2\sin{\theta}-1 \right) c_2\sin{\theta} +2r\left[2(c_1+c_2)-s_0\right] =0\;,\\
\frac{\partial E }{\partial c_2}&= \left(c_1c_2\sin{\theta}-1 \right) c_1\sin{\theta} +2r\left[2(c_1+c_2)-s_0\right] =0\;,\\
\frac{\partial E}{\partial \theta}&= \left(c_1c_2\sin{\theta}-1 \right)c_1c_2\cos{\theta}=0\;.
\end{aligned}    
\label{eq:lineareqssquare}
\end{equation}
}

{Clearly the compatible state $a=1$ and $p=s_0$ identically satisfies all three equations. This solution requires
\begin{equation}
\begin{aligned}
&c_1c_2\sin\theta=1\;,\\
&2(c_1+c_2)=s_0\;,
\end{aligned}
\end{equation}
with solution
\begin{equation}
\begin{aligned}
&c_1=\frac{s_0}{4}+\frac{1}{4}\sqrt{p_0^2-\frac{s_0^{*2}}{\sin\theta}}\;,\\
&c_2=\frac{s_0}{4}-\frac{1}{4}\sqrt{p_0^2-\frac{s_0^{*2}}{\sin\theta}}\;,
\end{aligned}
\end{equation}
provided
\begin{equation}
\sin\theta\geq(s_0^*/s_0)^2\;.
\label{eq:theta}
\end{equation}
or $s_0>s_0^*/\sqrt{\sin\theta}$.
In other words for any value of $s_0>s_0^*$ the compatible solution is a family of quadrilaterals with $a=1$, $p=p_0$ and  tilt angle $\theta$ varying in the range specified by Eq.~\eqref{eq:theta}. At $s_0=s_0^*$ there is a single solution corresponding to a square with $\theta=\pi/2$ and $c_1=c_2=1$.}

{When $\theta=\pi/2$, the last of equations ~\eqref{eq:lineareqssquare} is identically satisfied. For $s_0<s_0^*$ there is then a state with $c_1=c_2=c$ given by the solution of
\begin{equation}
c^3+\left(8 r-1\right)c- 2 s_0^* r =0\;.  
\label{eq:cub}
\end{equation}
This is the incompatible regime. There is a single ground state corresponding to a regular square and the energy is gapped.
If $r\gg1$, corresponding to the case where perimeter deformations are much more costly than area deformation, Eq.~\eqref{eq:cub} has solution $c\simeq s_0/4$, corresponding to $p\simeq s_0$ and $a\simeq s_0^2/16$, with $p/\sqrt{a}\simeq 4=s_0^*$.
In the opposite limit of $r\ll 1$ we find $c\simeq 1$, corresponding to $a\simeq 1$ and $p\simeq 4$, with $p/\sqrt{a}\simeq 4=s_0^*$.
In general the compatible cell has $p>s_0$ and $a<1$, with $p/\sqrt{a} =s_0^*$ for all $s_0\leq s_0^*$.} 

\begin{figure}
\centering
\includegraphics[width=\linewidth]{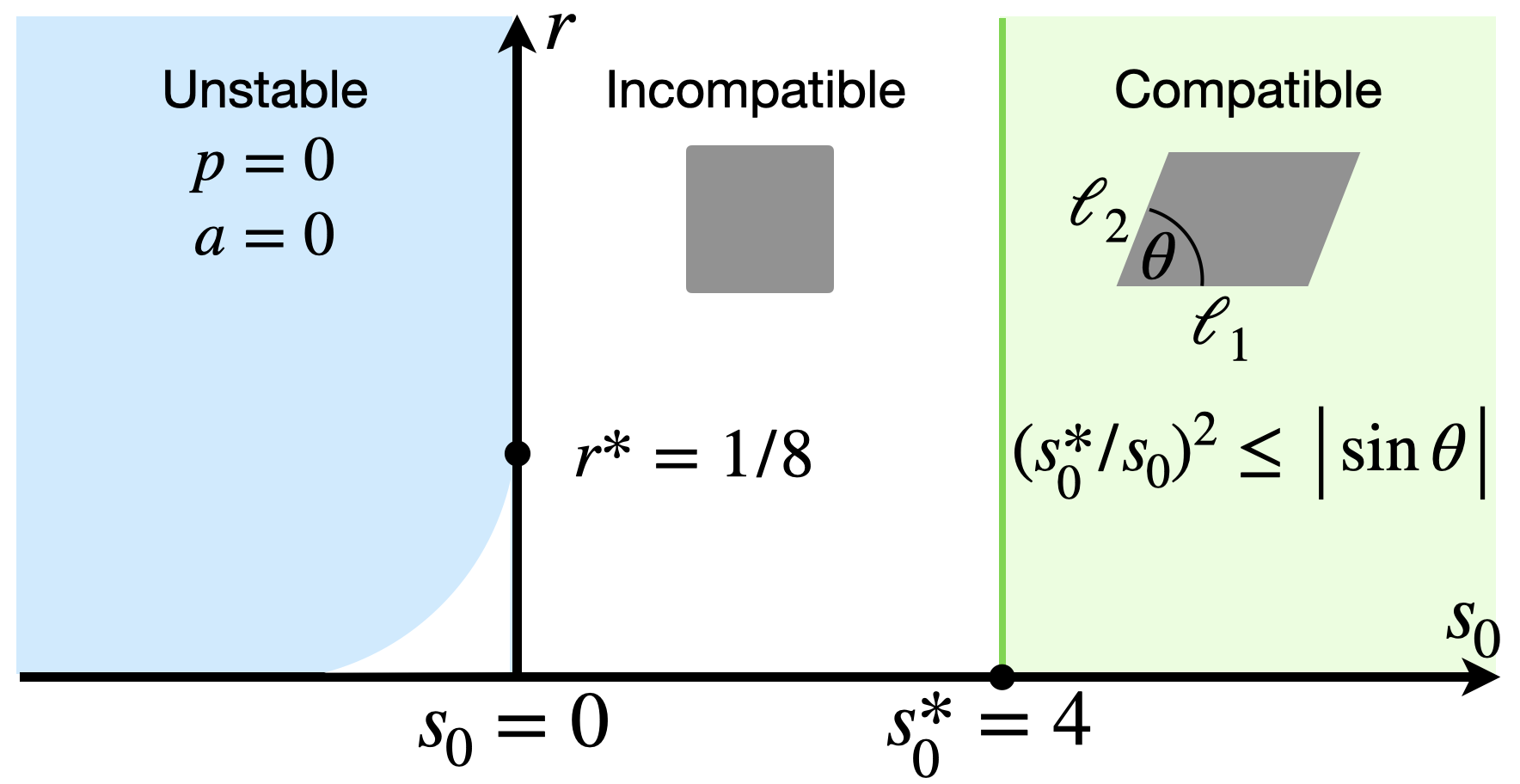}
\caption{Mean-field phase diagram for a $4$-sided VM in the $(s_0,r)$ plane. For $s_0>s_0^*=4$ the ground state is compatible, corresponding to a family of zero-energy quadrilaterals parametrized by the tilt angle $\theta$. At the critical point $s_0=s_0^*$ there is a unique ground state corresponding to a square cell with $a=1$ and $p=4$. For $s_0<s_0^*$ the system cannot satisfy both area and perimeter constraints and the ground state is a square with side determined by the real solution of the cubic equation \eqref{eq:cub}. The incompatible regime extends into the region $s_0<0$ for $r<r^*$. The blue region corresponds to a collapsed cell with $a=p=0$.}
\label{fig:square}
\end{figure}

The solution $c=0$ corresponds to a collapsed cell with $a=p=0$ and minimum energy $E_m=(1+rs_0^2)/2$.
Imposing that $E(c)>E_m$, 
where $c$ is the solution of Eq.~\eqref{eq:cub}, yields the constraint $c>c_m$, with $c_m=6rs_0/(2s_0^*r-1)>0$. 
 Using this condition it can be shown that, as demonstrated in Ref.~\cite{farhadifar2007influence}, the system is unstable, corresponding to a collapsed cell, for $r\leq 1/8$ and $s_0<-\frac{2}{r}\left(\frac{1-8r}{6}\right)^{3/2}$. 
 The corresponding phase diagram is shown in Fig.~\ref{fig:square}.

\begin{figure}
	\centering
	\includegraphics[width=\linewidth]{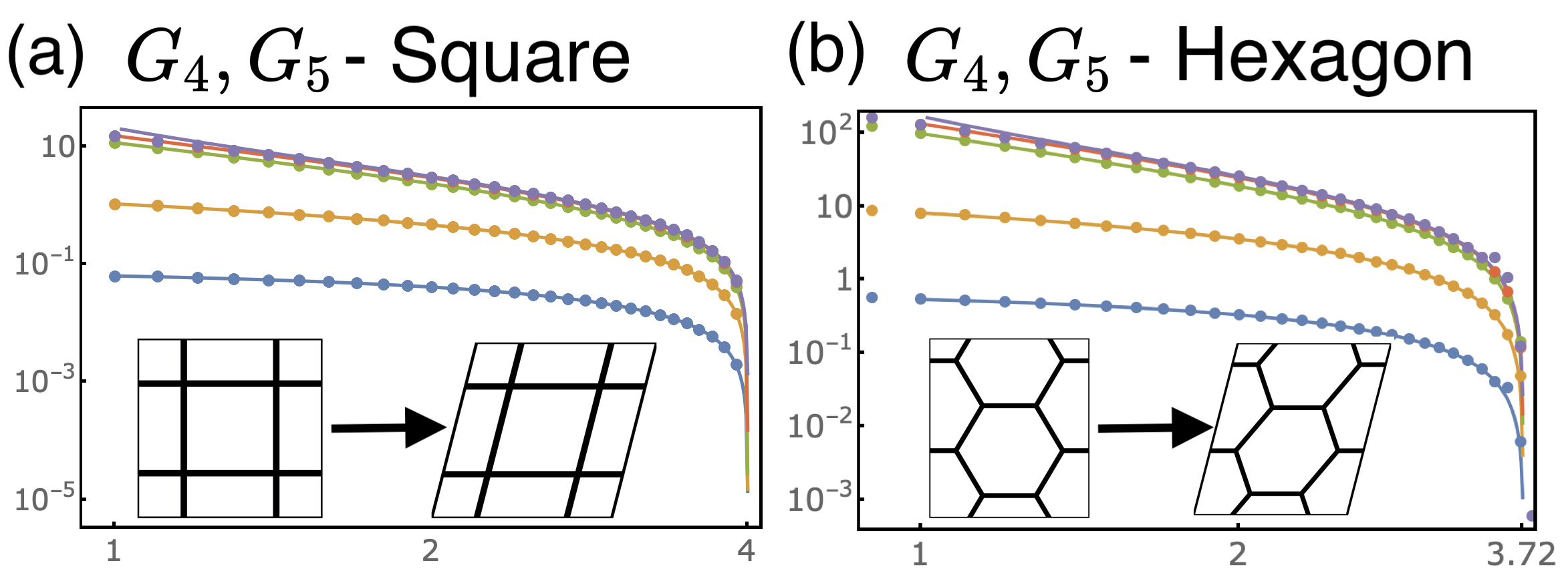}
	\caption{Extension of  \figref{fig:Moduli}  to square and hexagonal tissue model: Comparison of analytical (solid) and numerical (points) elastic moduli of an (a) $n=4$ and (b) $n=6$ VM as functions of target shape parameter $s_0$ on a log-log scale. The compatible/incompatible transition is at $s_0^*=4$ and $s_0^*=192^{1/4}\approx 3.72$ respectively.} 
	\label{fig:app_Moduli}
\end{figure}

In general, we can calculate the incompatible ground state  for any $n$-sided  polygonal cell by noting that in this regime the ground-state metric is isotropic and can be written as
 \begin{equation}
     \g_0 = c^2 \begin{pmatrix}  1 & 0 \\  0 & 1 \end{pmatrix}
     \label{eq:gsMet}
 \end{equation}
 with $c$ to be determined by energy minimization.  Upon substituting $\g=\go$ in Eq.~\eqref{eq:Energy}, the energy $E_n$ of an $n$-side cell reads
  \begin{equation}
  \begin{split}
      E_3 &= \frac{1}{2} \left(\frac{\sqrt{3} }{4} c^2 - 1\right)^2+\frac{1}{2} r \left(3 c-p_0\right)^2\;,\\
      E_4 &= \frac{1}{2} \left(c^2-1\right)^2+\frac{1}{2} r \left(4 c-p_0\right)^2\;,\\
      E_6 &= \frac{1}{2} \left(\frac{3 \sqrt{3}}{2}  c^2-1\right)^2+\frac{1}{2} r \left(6 c-p_0\right)^2\;.
  \end{split}
 \end{equation}
The value of $c$ that minimizes the energy  is the solution of a cubic equation given in Eq.~\eqref{eq:cub} for $n=4$ and by the following equations for $n=3,6$ 
  \begin{equation}
  \begin{split}
       c^3+c \left(96 r-\frac{4}{\sqrt{3}}\right)-32 p_0 r = 0\;,~~~n=3\;,\\
      9 c^3+c \left(24 r-2 \sqrt{3}\right)-4 p_0 r = 0\;,~~~n=6\;.
  \end{split}
  \label{eq:gsC}
 \end{equation}
The mechanical response to small perturbations relative to the ground state for a triangular VM is shown in \figref{fig:Moduli} where analytical and numerical results are compared and are in very good agreement. In \figref{fig:app_Moduli} we compare numerical and analytical calculations of the shear modulus for square and hexagonal tissue VM. 
The elastic tensor in the hexagonal case is of the form \eqref{eq:ET} with
\begin{equation}
\label{eq:lame_hex}
	\begin{split}
		\lambda  &=  \frac{9 r}{2 c}\left(  3  s_0 - 14 c \right)+ 3 \sqrt{3} \;,\\
		\mu  &= \frac{9 r}{2 c}  (6 c -   s_0 )\;.
	\end{split}
\end{equation}
Detailed derivations of this and other results as well as comparison with numerical simulations can be found in an attached Mathematica Notebook.


\bibliographystyle{unsrt}
\bibliography{references}

\begin{thebibliography}{10}

\bibitem{gomez2020measuring}
Manuel G{\'o}mez-Gonz{\'a}lez, Ernest Latorre, Marino Arroyo, and Xavier
  Trepat.
\newblock Measuring mechanical stress in living tissues.
\newblock {\em Nature Reviews Physics}, 2(6):300--317, 2020.

\bibitem{angelini2011glass}
Thomas~E Angelini, Edouard Hannezo, Xavier Trepat, Manuel Marquez, Jeffrey~J
  Fredberg, and David~A Weitz.
\newblock Glass-like dynamics of collective cell migration.
\newblock {\em Proceedings of the National Academy of Sciences},
  108(12):4714--4719, 2011.

\bibitem{puliafito2012collective}
Alberto Puliafito, Lars Hufnagel, Pierre Neveu, Sebastian Streichan, Alex
  Sigal, D~Kuchnir Fygenson, and Boris~I Shraiman.
\newblock Collective and single cell behavior in epithelial contact inhibition.
\newblock {\em Proceedings of the National Academy of Sciences},
  109(3):739--744, 2012.

\bibitem{sadati2013collective}
Monirosadat Sadati, Nader~Taheri Qazvini, Ramaswamy Krishnan, Chan~Young Park,
  and Jeffrey~J Fredberg.
\newblock Collective migration and cell jamming.
\newblock {\em Differentiation}, 86(3):121--125, 2013.

\bibitem{park2015unjamming}
Jin-Ah Park, Jae~Hun Kim, Dapeng Bi, Jennifer~A Mitchel, Nader~Taheri Qazvini,
  Kelan Tantisira, Chan~Young Park, Maureen McGill, Sae-Hoon Kim, Bomi Gweon,
  et~al.
\newblock Unjamming and cell shape in the asthmatic airway epithelium.
\newblock {\em Nature materials}, 14(10):1040--1048, 2015.

\bibitem{mongera2018fluid}
Alessandro Mongera, Payam Rowghanian, Hannah~J Gustafson, Elijah Shelton,
  David~A Kealhofer, Emmet~K Carn, Friedhelm Serwane, Adam~A Lucio, James
  Giammona, and Otger Camp{\`a}s.
\newblock A fluid-to-solid jamming transition underlies vertebrate body axis
  elongation.
\newblock {\em Nature}, 561(7723):401--405, 2018.

\bibitem{bi2015density}
Dapeng Bi, JH~Lopez, Jennifer~M Schwarz, and M~Lisa Manning.
\newblock A density-independent rigidity transition in biological tissues.
\newblock {\em Nature Physics}, 11(12):1074--1079, 2015.

\bibitem{bi2016motility}
Dapeng Bi, Xingbo Yang, M~Cristina Marchetti, and M~Lisa Manning.
\newblock Motility-driven glass and jamming transitions in biological tissues.
\newblock {\em Physical Review X}, 6(2):021011, 2016.

\bibitem{krajnc2020solid}
Matej Krajnc.
\newblock Solid--fluid transition and cell sorting in epithelia with junctional
  tension fluctuations.
\newblock {\em Soft Matter}, 16(13):3209--3215, 2020.

\bibitem{krajnc2021active}
Matej Krajnc, Tomer Stern, and Clement Zankoc.
\newblock Active instability of cell-cell junctions at the onset of tissue
  fluidity.
\newblock {\em arXiv preprint arXiv:2101.07058}, 2021.

\bibitem{storm2005nonlinear}
Cornelis Storm, Jennifer~J Pastore, Fred~C MacKintosh, Tom~C Lubensky, and
  Paul~A Janmey.
\newblock Nonlinear elasticity in biological gels.
\newblock {\em Nature}, 435(7039):191--194, 2005.

\bibitem{chen2018branches}
Bryan Gin-ge Chen and Christian~D Santangelo.
\newblock Branches of triangulated origami near the unfolded state.
\newblock {\em Physical Review X}, 8(1):011034, 2018.

\bibitem{moshe2018geometric}
Michael Moshe, Mark~J Bowick, and M~Cristina Marchetti.
\newblock Geometric frustration and solid-solid transitions in model 2d tissue.
\newblock {\em Physical review letters}, 120(26):268105, 2018.

\bibitem{sahu2019nonlinear}
Preeti Sahu, Janice Kang, Gonca Erdemci-Tandogan, and M~Lisa Manning.
\newblock Nonlinear analysis of the fluid-solid transition in a model for
  ordered biological tissues.
\newblock {\em arXiv preprint arXiv:1905.12714}, 2019.

\bibitem{damavandi2021energeticI}
Ojan~Khatib Damavandi, Varda~F Hagh, Christian~D Santangelo, and M~Lisa
  Manning.
\newblock Energetic rigidity: a unifying theory of mechanical stability.
\newblock {\em arXiv preprint arXiv:2102.11310}, 2021.

\bibitem{damavandi2021energeticII}
Ojan~Khatib Damavandi, Varda~F Hagh, Christian~D Santangelo, and M~Lisa
  Manning.
\newblock Energetic rigidity ii: Applications in examples of biological and
  underconstrained materials.
\newblock {\em arXiv preprint arXiv:2107.06868}, 2021.

\bibitem{tong2021linear}
Sijie Tong, Navreeta~K Singh, Rastko Sknepnek, and Andrej Kosmrlj.
\newblock Linear viscoelastic properties of the vertex model for epithelial
  tissues.
\newblock {\em arXiv preprint arXiv:2102.11181}, 2021.

\bibitem{honda1983geometrical}
Hisao Honda.
\newblock Geometrical models for cells in tissues.
\newblock {\em International review of cytology}, 81:191--248, 1983.

\bibitem{nagai2001dynamic}
Tatsuzo Nagai and Hisao Honda.
\newblock A dynamic cell model for the formation of epithelial tissues.
\newblock {\em Philosophical Magazine B}, 81(7):699--719, 2001.

\bibitem{staple2010mechanics}
Douglas~B Staple, Reza Farhadifar, J-C R{\"o}per, Benoit Aigouy, Suzanne Eaton,
  and Frank J{\"u}licher.
\newblock Mechanics and remodelling of cell packings in epithelia.
\newblock {\em The European Physical Journal E}, 33(2):117--127, 2010.

\bibitem{farhadifar2007influence}
Reza Farhadifar, Jens-Christian R{\"o}per, Benoit Aigouy, Suzanne Eaton, and
  Frank J{\"u}licher.
\newblock The influence of cell mechanics, cell-cell interactions, and
  proliferation on epithelial packing.
\newblock {\em Current Biology}, 17(24):2095--2104, 2007.

\bibitem{chiou2012mechanical}
Kevin~K Chiou, Lars Hufnagel, and Boris~I Shraiman.
\newblock Mechanical stress inference for two dimensional cell arrays.
\newblock {\em PLoS computational biology}, 8(5):e1002512, 2012.

\bibitem{fletcher2014vertex}
Alexander~G Fletcher, Miriam Osterfield, Ruth~E Baker, and Stanislav~Y
  Shvartsman.
\newblock Vertex models of epithelial morphogenesis.
\newblock {\em Biophysical journal}, 106(11):2291--2304, 2014.

\bibitem{alt2017vertex}
Silvanus Alt, Poulami Ganguly, and Guillaume Salbreux.
\newblock Vertex models: from cell mechanics to tissue morphogenesis.
\newblock {\em Philosophical Transactions of the Royal Society B: Biological
  Sciences}, 372(1720):20150520, 2017.

\bibitem{barton2017active}
Daniel~L Barton, Silke Henkes, Cornelis~J Weijer, and Rastko Sknepnek.
\newblock Active vertex model for cell-resolution description of epithelial
  tissue mechanics.
\newblock {\em PLoS computational biology}, 13(6):e1005569, 2017.

\bibitem{merkel2017triangles}
Matthias Merkel, Rapha{\"e}l Etournay, Marko Popovi{\'c}, Guillaume Salbreux,
  Suzanne Eaton, and Frank J{\"u}licher.
\newblock Triangles bridge the scales: Quantifying cellular contributions to
  tissue deformation.
\newblock {\em Physical Review E}, 95(3):032401, 2017.

\bibitem{merkel2018geometrically}
Matthias Merkel and M~Lisa Manning.
\newblock A geometrically controlled rigidity transition in a model for
  confluent 3d tissues.
\newblock {\em New Journal of Physics}, 20(2):022002, 2018.

\bibitem{popovic2021inferring}
Marko Popovi{\'c}, Valentin Druelle, Natalie~A Dye, Frank J{\"u}licher, and
  Matthieu Wyart.
\newblock Inferring the flow properties of epithelial tissues from their
  geometry.
\newblock {\em New Journal of Physics}, 23(3):033004, 2021.

\bibitem{grossman2021instabilities}
Doron Grossman and Jean-Francois Joanny.
\newblock Instabilities and geometry of growing tissue.
\newblock {\em arXiv preprint arXiv:2108.05326}, 2021.

\bibitem{huang2021shear}
Junxiang Huang, James Cochran, Suzanne~M Fielding, M.~Cristina Marchetti, and
  Dapeng Bi.
\newblock Shear-driven solidification and non-linear elasticity in epithelial
  tissues.
\newblock {\em arXiv preprint arXiv:2109.10374}, 2021.

\bibitem{seung1988defects}
Hyunjune~Sebastian Seung and David~R Nelson.
\newblock Defects in flexible membranes with crystalline order.
\newblock {\em Physical Review A}, 38(2):1005, 1988.

\bibitem{lubensky2015phonons}
TC~Lubensky, CL~Kane, Xiaoming Mao, Anton Souslov, and Kai Sun.
\newblock Phonons and elasticity in critically coordinated lattices.
\newblock {\em Reports on Progress in Physics}, 78(7):073901, 2015.

\bibitem{sheinman2012nonlinear}
M~Sheinman, CP~Broedersz, and FC~MacKintosh.
\newblock Nonlinear effective-medium theory of disordered spring networks.
\newblock {\em Physical Review E}, 85(2):021801, 2012.

\bibitem{yan2019multicellular}
Le~Yan and Dapeng Bi.
\newblock Multicellular rosettes drive fluid-solid transition in epithelial
  tissues.
\newblock {\em Physical Review X}, 9(1):011029, 2019.

\bibitem{siefert2021euclidean}
Emmanuel Si{\'e}fert, Ido Levin, and Eran Sharon.
\newblock Euclidean frustrated ribbons.
\newblock {\em Physical Review X}, 11(1):011062, 2021.

\bibitem{kupferman2020continuum}
Raz Kupferman, Ben Maman, and Michael Moshe.
\newblock Continuum mechanics of a cellular tissue model.
\newblock {\em Journal of the Mechanics and Physics of Solids}, 143:104085,
  2020.

\bibitem{scheibner2020odd}
Colin Scheibner, Anton Souslov, Debarghya Banerjee, Piotr Sur{\'o}wka,
  William~TM Irvine, and Vincenzo Vitelli.
\newblock Odd elasticity.
\newblock {\em Nature Physics}, 16(4):475--480, 2020.

\bibitem{kupferman2018variational}
Raz Kupferman and Cy~Maor.
\newblock Variational convergence of discrete geometrically-incompatible
  elastic models.
\newblock {\em Calculus of Variations and Partial Differential Equations},
  57(2):1--27, 2018.

\bibitem{brakke1992surface}
Kenneth~A Brakke.
\newblock The surface evolver.
\newblock {\em Experimental mathematics}, 1(2):141--165, 1992.

\end{thebibliography}


\end{document}